\documentclass[12pt,a4paper]{article}
\usepackage{graphicx}
\usepackage{amsmath}
\usepackage{amssymb}
\usepackage{hyperref}
\usepackage[height=8.8in,width=6.45in]{geometry}

\numberwithin{equation}{section}

\usepackage{tikz}
\usetikzlibrary{arrows}
\tikzset{>=latex}

\def\bx{\mathbf{x}}
\def\fg{\mathfrak{g}}
\def\cN{\mathcal{N}{}}

\def\tr{\mathop{\mathrm{tr}}\nolimits}
\def\vol{\mathop{\mathrm{vol}}\nolimits}
\def\SU{\mathop{\mathrm{SU}}\nolimits}
\def\Ind{\mathop{\mathrm{Ind}}\nolimits}
\newcommand{\cA}{\mathcal{A}\,}
\newcommand{\cH}{\mathcal{H}}
\newcommand{\bR}{\mathbb{R}}
\newcommand{\bC}{\mathbb{C}}
\def\braket#1#2{\langle#1|#2\rangle}
\def\bra#1{\langle#1|}
\def\ket#1{|#1\rangle}
\def\ut{\underline{t}}
\def\lambda{\mathcal{R}}
\def\emphasize#1{\\ \centerline{\textcolor{red}{\emph{#1}}} \\ }
\begin{document}

\begin{titlepage}

\begin{flushright}
IPMU 12-0141 \\
UT-12-24
\end{flushright}
\vskip 2cm

\begin{center}
{\Large \bfseries
4d partition function on $S^1\times S^3$ \\[1em]
and 2d Yang-Mills with nonzero area
}

\vskip 1.2cm

Yuji Tachikawa$^{\sharp,\flat}$

\bigskip
\bigskip

\begin{tabular}{ll}
$^\flat$  & Department of Physics, Faculty of Science, \\
& University of Tokyo,  Bunkyo-ku, Tokyo 133-0022, Japan\\
$^\sharp$  & Kavli Institute for the Physics and Mathematics of the Universe, \\
& University of Tokyo,  Kashiwa, Chiba 277-8583, Japan
\end{tabular}

\vskip 1.5cm

\textbf{abstract}
\end{center}

\medskip
\noindent

We argue that 6d $\cN{=}(2,0)$ theory on $S^1\times S^3\times C_2$ reduces to the 2d q-deformed Yang-Mills on $C_2$ at finite area, as a small extension to the result of Gadde, Rastelli, Razamat and Yan. This is done by computing the partition function on $S^1\times S^3$ of 4d $\cN{=}2$ supersymmetric non-linear sigma model on $T^*G_\bC$, which gives the propagator of the 2d Yang-Mills. 

\bigskip
\vfill
\end{titlepage}

\tableofcontents

\section{Introduction and Summary}

In \cite{Gadde:2009kb,Gadde:2011ik,Gadde:2011uv}, a remarkable observation was made that the superconformal index of the 4d $\cN{=}2$ theory associated to a Riemann surface $C_2$ by Gaiotto in \cite{Gaiotto:2009we} is equal to the partition function of 2d q-deformed Yang-Mills on the same Riemann surface $C_2$ in the \emph{zero area limit}, or a generalization thereof.
As discussed in \cite{Gaiotto:2011xs}, the 4d $\cN{=}2$ theory associated to $C_2$ by Gaiotto  in \cite{Gaiotto:2009we} should be thought of as the compactification of 6d $\cN{=}(2,0)$ theory on the Riemann surface $C_2$ in the \emph{zero area limit}, in order to decouple various Kaluza-Klein modes so that we have a genuine 4d theory.

The aim of this short note is to argue that the partition function on $S^1\times S^3$ of the compactification of 6d $\cN{=}(2,0)$ theory on the Riemann surface $C_2$ with nonzero area $\cA$, is equal to the partition function of 2d q-deformed Yang-Mills on $C_2$ with nonzero area $\cA$.\footnote{Before proceeding, it would be useful to recall recent advances in  the localization in general. First, there is the localization down to matrix models, of 2d theory on $S^2$ \cite{Benini:2012ui,Doroud:2012xw}, of 3d theory on $S^3$  \cite{Kapustin:2009kz}, of 4d theory on $S^4$  \cite{Pestun:2007rz},  and of 5d theory on $S^5$ \cite{Kallen:2012cs,Hosomichi:2012ek,Kallen:2012va,Kim:2012av,Kallen:2012zn}.   The moral is that, as far as BPS quantities are concerned, 
\emphasize{$N$-dimensional  theory on $S^N$ can be localized  to a $0$-dimensional field theory,} i.e.~a matrix model.  
Using these localization techniques, it was then found that the partition function on $S^4$ of 4d $\cN{=}2$ theory associated to a Riemann surface $C_2$ is equal to the partition function on $C_2$ of 2d Toda theory \cite{Alday:2009aq,Wyllard:2009hg}, and also that the partition function on $S^3$ of 3d $\cN{=}2$ theory associated to a hyperbolic manifold $H_3$ is equal to the partition function on $H_3$ of Chern-Simons theory \cite{Dimofte:2011ju,Terashima:2011xe,Dimofte:2011py}. 
In other words, 6d $\cN{=}(2,0)$ theory becomes 2d Toda theory or 3d Chern-Simons theory if reduced on $S^4$ or $S^3$, respectively.
Although these facts were observed by computing the partition functions by localizing down to zero dimensions, the moral should be that, as far as BPS quantities are concerned, \emphasize{$(N+n)$-dimensional theory on $S^N$  can be localized to an $n$-dimensional field theory.}
Note that the various matrix models above are to be thought of as a special case when $n=0$. 
Also, the lore that the BPS sector of 6d $\cN{=}(2,0)$ theory on $S^1$ is wholly captured by maximally supersymmetric Yang-Mills in 5d is again a special case when $N=1$ and $n=5$. 
Very recently, 5d minimally supersymmetric Yang-Mills on $S^3\times \bR^2$ was analyzed \cite{Kawano:2012up}, in which it was shown that it localizes to 2d Yang-Mills on $\bR^2$. The localization on $S^4\times S^1$ was also done in \cite{Kim:2012gu,Terashima:2012ra}.
Furthermore, the manifold on which the theory is localized does not have to be a completely round $S^N$; any manifold which admits rigid supersymmetry in the sense of \cite{Festuccia:2011ws} should be usable. For example, localization on squashed spheres was also performed in \cite{Hama:2010av,Hama:2011ea,Imamura:2011wg,Hama:2012bg}. Localization on a finite quotient of $S^3$ was done in \cite{Benini:2011nc}.  }

\begin{figure}[h!]
\[
\def\X{(4,-2.2)}
\def\Y{(-4,-2.2)}
\def\Z{(0,-1.7)}
\begin{tikzpicture}
\node (132) at (0,0) {6d $\cN{=}(2,0)$ theory on $S^1\times S^3 \times C_2$};
\path (132) to ++\X node (32){5d SYM on $S^3\times C_2$ } to ++\Y node(2){2d qYM on $C_2$} to ++\Z node(Z){$Z$};
\path (132) to ++\Y node (12){CS on $S^1\times C_2$ } to ++\Z node(1){some QM on $S^1$};
\path (132) to ++\Z node (13){4d $\cN{=}2$ theory on $S^1\times S^3$ } to ++\X node(3){3d $\cN{=}4$  theory on $S^3$};
\draw[->] (132) --(32);
\draw[->] (132) --(13);
\draw[->] (132) --(12);
\draw[->] (32) --(2);
\draw[->] (32) --(3);
\draw[->] (12) --(1);
\draw[->] (12) --(2);
\draw[->] (13) --(1);
\draw[->] (13) --(3);
\draw[->] (1) --(Z);
\draw[->] (2) --(Z);
\draw[->] (3) --(Z);
\draw[<->,thick,dotted,red
%decorate,decoration={coil,pre length=8pt,post length=5pt}
] (13) to (2);
\draw[<->,thick,dashed,green!80!black%,
%decorate,decoration={saw,pre length=8pt,post length=5pt}
] (12)  to (3);
\draw[<->,thick,dash pattern=on 10pt off 2pt,blue
%decorate,decoration={zigzag,pre length=8pt,post length=5pt}
] (32)  to (1);
\end{tikzpicture}
\]
\caption{Interrelation of various theories, starting from 6d $\cN{=}(2,0)$ on $S^1\times S^3\times C_2$. The black arrows show the dimensional reductions. The red, green and blue double arrows show the manifestations of various dualities discussed in the literature, see the main text. \label{figure}}
\end{figure}
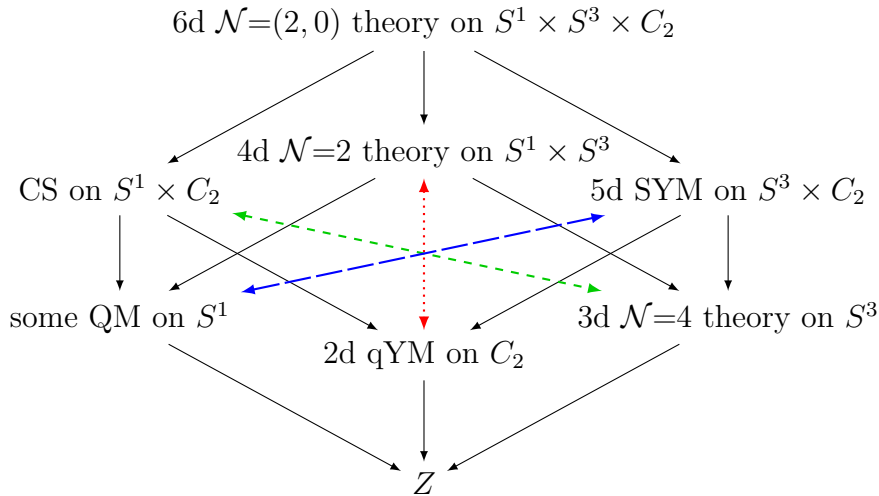

We are interested in 6d $\cN{=}(2,0)$ theory on $S^1\times S^3\times C_2$. This system can be analyzed in many ways, according to the different orders of piecemeal reductions, as shown in Figure~\ref{figure}. The relation (given by a \textcolor{red}{red} dotted arrow) between 4d $\cN{=}2$ theory on $S^1\times S^3$ and 2d q-deformed Yang-Mills on $C_2$ is the one discovered in \cite{Gadde:2011ik}; the relation (given by a \textcolor{green!80!black}{green} dashed arrow) between Chern-Simons theory on $S^1\times C_2$ and 3d $\cN{=}4$ theory on $S^3$ is a version of 3+3 duality \cite{Dimofte:2011ju,Terashima:2011xe,Dimofte:2011py}; the relation (given by a \textcolor{blue}{blue} broken arrow) says that the partition function of 5d super Yang-Mills on a five-manifold automatically has the structure of an index over a vector space; this was recently checked when the five-manifold is $S^5$ \cite{Kim:2012av}.

The most solid way to see that 6d $\cN{=}(2,0)$ theory on $S^1\times S^3\times C_2$  reduces to 2d q-deformed Yang-Mills with nonzero area would be to reduce it first to 5d maximally supersymmetric Yang-Mills on $S^3\times C_2$, and then to reduce it further on $S^3$ to get a 2d theory on $C_2$, as an extension of a recent work \cite{Kawano:2012up}. In this paper, we instead take a rather round-about approach, to study the partition function on $S^1\times S^3$ of 4d $\cN{=}2$ theory obtained by compactifying 6d $\cN{=}(2,0)$ theory on $C_2$ with finite area, studied in \cite{Gaiotto:2011xs}.

Recall that in \cite{Gadde:2011ik} the relationship between 4d $\cN{=}2$ theory and 2d q-deformed Yang-Mills was made by performing the pants decomposition of the Riemann surface $C_2$. The only place where the area mattered is at the treatment of the `propagator', i.e.~the partition function on the cylinder $S^1\times$ a finite segment of 2d q-deformed Yang-Mills. 
Hence, to include the effect of having nonzero area, we only need to study the 4d theory one obtains by putting 6d $\cN{=}(2,0)$ theory on a cylinder with area $\cA$, and to show its partition function on $S^1\times S^3$ agrees with the propagator of 2d q-deformed Yang-Mills for area $\cA$. 

In \cite{Gaiotto:2008sa,Gaiotto:2011xs}, it was shown that  6d $\cN{=}(2,0)$ theory of type $G$ on a cylinder with $\cA$ is essentially a 4d $\cN{=}2$ sigma model whose target space is 
a hyperk\"ahler manifold $T^*G_\bC$, whose metric has an overall factor of $\cA^{-1}$: \begin{equation}
ds^2_\cA=\frac{1}{\cA} ds^2_{\cA=1}.\label{metric}
\end{equation} We denote the sigma model by $X_{\cA}$. This sigma model has $G_L\times G_R$ flavor symmetry. 
We put this theory on $S^1\times S^3$, introduce exponentiated chemical potentials $t,y,v$ associated to the spacetime and R-symmetries as in \cite{Gadde:2009kb}.  
We mostly set $y=1$, $v=t$  as in \cite{Gadde:2011ik}, so that the corresponding 2d theory becomes the q-deformed Yang-Mills, where $q=t^3$. We then want to calculate the partition function of the theory $X_{\cA}$, which we denote by $Z_\cA(q,\bx_L,\bx_R)$, where  $\bx_L,\bx_R$ are fugacities for flavor symmetries $G_L\times G_R$.
Our objective is to show that this $Z_\cA$ is the propagator of the 2d Yang-Mills.

The rest of the note is organized as follows. In Sec.~\ref{calc} we first find the partition function heuristically, using the relation among $X_{\cA}$, $X_{\cA'}$ and $X_{\cA+\cA'}$.
In Sec.~\ref{check}, we give a check of this result by studying the decomposition of the Hilbert space under $G_L\times G_R$.
In Sec.~\ref{generalization}, we speculate how to relax the conditions $y=1$, $v=t$.
We assume the reader is familiar with the results of and the notations in \cite{Gadde:2011ik,Gadde:2011uv,Gaiotto:2011xs}.

\section{Partition function of the non-linear sigma model}\label{calc}
Let us take two copies of the theory with different area, $X_{\cA}$ and $X_{\cA'}$, and couple an $\cN{=}2$ vector multiplet $V_G$ of gauge group $G$ to the diagonal subgroup of $G_R$ of $X_\cA$ and $G_L$ of $X_{\cA'}$. 
As shown in \cite{Gaiotto:2008sa,Gaiotto:2011xs},  when the multiplet $V_G$ has zero kinetic term and is auxiliary, the resulting theory is just $X_{\cA+\cA'}$:\footnote{Let us illustrate this property with a toy example with no supersymmetry and $G=\mathrm{U}(1)$. The theory on the left hand side has the Lagrangian $
L=\frac{1}{\cA}(\partial_\mu\phi-A_\mu)^2 +\frac{1}{\cA'}(\partial_\mu\phi'+A_\mu)^2
$ where the periodicity of both $\phi$ and $\phi'$ is $2\pi$. Fix the gauge by demanding $\phi'=0$.  Then $A_\mu$ can be solved in terms of $\partial_\mu \phi$. Plugging it back in, one obtains $
L=\frac{1}{\cA+\cA'} (\partial_\mu\phi) ^2.
$ This analysis can be easily $\cN{=}2$-supersymmetrized. } \begin{equation}
X_{\cA} + V_G + X_{\cA'} = X_{\cA+\cA'}.
\end{equation} As a statement of the partition function, this becomes \begin{equation}
\int [d\bx] Z_\cA(q,\bx_L,\bx)\eta(q,\bx)Z_{\cA'}(q,\bx,\bx_R) = Z_{\cA+\cA'}(q,\bx_L,\bx_R)\label{composition}
\end{equation} where \begin{equation}
\eta(q,\bx)=\exp\left[\sum_{n=1}^\infty \frac{-2q^{n}}{1-q^{n}}\chi_\text{adj}(\bx^n)\right]\label{gauge}
\end{equation} is the contribution to the index from a gauge multiplet of gauge group $G$, and $[d\bx]$ is the Haar measure.

Topologically, the target manifold has the form \begin{equation}
T^*G_\bC\simeq G\times (\fg\oplus\fg\oplus\fg) \label{tg}
\end{equation}
where the zero-section $G$ is preserved by $\SU(2)_R$, while $\fg\oplus\fg\oplus\fg$ transforms as a triplet.
This means that the part $G$ remains massless even on $S^1\times S^3$ and that it becomes (part of) the zero modes.  
Let us assume then that the partition function  has the factorized structure \begin{equation}
Z_{\cA}(q,\bx_L,\bx_R)= Z^\text{1-loop}(q,\bx_L,\bx_R) Z^\text{zero mode}_\cA(q,\bx_L,\bx_R).\label{full}
\end{equation} Note that the $\cA$ dependence only comes from the zero-mode integral. Otherwise the partition function on $S^1\times S^3$ is independent of parameters in the Lagrangian as shown in \cite{Festuccia:2011ws}.

The zero mode part is essentially a quantum mechanics on $G$, with a quadratic Hamiltonian coming from the metric \eqref{metric}, further reduced down on $S^3$: \begin{equation}
ds^2_{QM}=\frac{\vol S^3}{\cA} ds^2_{\cA=1}.
\end{equation} The theorem of Peter-Weyl says that  the Hilbert space $\cH$ of the quantum mechanics on $G$ decomposes as $
\cH=\bigoplus_\lambda \lambda\otimes \lambda^*
$ under $G\times G$,  where $\lambda$ runs over the irreducible representations of $G$.
The eigenvalues of the Laplacian under the metric \eqref{metric} is given by $ C_2(\lambda)\cA/\vol S^3$, where $C_2(\lambda)$ is the quadratic Casimir of the representation $\lambda$. Therefore \begin{equation}
\begin{aligned}
Z^\text{zero mode}_\cA(q,\bx_L,\bx_R) &= \sum_\lambda e^{-C_2(\lambda)\cA \beta /\vol S^3} \tr_{\lambda\otimes \lambda^*}  \bx_L \otimes \bx_R \\
&=\sum_\lambda \chi_\lambda(\bx_L)  e^{-C_2(\lambda)  \cA\beta /\vol S^3}\chi_\lambda^*(\bx_R) \\
&=\sum_\lambda \braket{\bx_L}{\lambda}e^{-C_2(\lambda)\cA\beta/\vol S^3}\braket{\lambda}{\bx_R}
\end{aligned}
\end{equation} where $\chi_\lambda(\bx)=\braket{\bx}{\lambda}$ is the character in the representation $\lambda$ of the exponential of the fugacity $\bx$. Note that this quantity is originally a trace, but looks like a propagator.

The one-loop term is uniquely determined by demanding that the full partition function \eqref{full}  satisfies the composition law \eqref{composition}. We conclude \begin{equation}
Z_{\cA}(q,\bx_L,\bx_R)=\eta(q,\bx_L)^{-1/2}\bra{\bx_L}\left[\sum_\lambda \ket{\lambda}e^{-C_2(\lambda)\cA\beta/\vol S^3}\bra{\lambda}\right]\ket{\bx_R}\eta(q,\bx_R)^{-1/2}.\label{answer}
\end{equation}
The part in the square brackets is the propagator of the 2d (q-deformed) Yang-Mills as is well known \cite{Cordes:1994fc}, and the factor $\eta(q,\bx)^{-1/2}$ is the necessary normalization factor as found in \cite{Gadde:2011ik}. This is what we wanted to demonstrate.

Recall that our system is based on 6d $\cN{=}(2,0)$ theory on $S^1\times S^3\times C_2$ 
where $C_2=\tilde S^1\times \text{(segment)}$. 
The first $S^1$ has the circumference $\beta$. 
Let us say the second $\tilde S^1$ has the circumference $\tilde \beta$ and the segment has the length $\ell$, so that $\cA=\tilde\beta \ell$. 
Then the exponent  $-C_2(\lambda)\beta\tilde\beta\ell/\vol S^3$ in \eqref{answer} is symmetric under the exchange of the circumferences $\beta$, $\tilde\beta$ of two $S^1$s.  Note also that the exponent is invariant under the simultaneous scaling \begin{equation}
\beta\to c\beta, \quad
\tilde\beta\to c\tilde\beta, \quad
 \ell \to c\ell,\quad 
 \vol S^3\to c^3\vol S^3.
\end{equation} This is in accord with the fact that the 6d theory is conformal.

Note also that in the description as 4d sigma model on $S^1\times S^3$, the fugacities $\bx_{L,R}$ are holonomies around $S^1$, while in the desciption as 2d sigma model on $\tilde S^1\times \text{(segment)}$ the same fugacities $\bx_{L,R}$ are holonomies around $\tilde S^1$. This comes from the fact that in 6d theory, one can have a \emph{holonomy around the surface} $S^1\times \tilde S^1$, which becomes an ordinary holonomy around $S^1$ when the 6d theory is reduced along $\tilde S^1$ or vice versa. 

\section{Decomposition under the action of $G_L\times G_R$}\label{check}
Let us perform a small check of the result by studying the decomposition of the Hilbert space under $G\times G$.  Essentially, we are doing  the Kaluza-Klein expansion of the sigma model with target $T^*G_\bC$ on $S^3$. The modes with nonzero momenta will have energy of order $\sim 1/r$ and the wavefunction of the zero modes will have  energy of order $\sim \cA/r^3$, where $r$ is the radius of $S^3$. Therefore, when $\cA/r^2$ is parametrically small, there is a separation of scales, allowing the use of the Born-Oppenheimer approximation where the non-zero modes as the fast modes, and the zero modes as the slow modes. 

The non-zero modes are essentially a free hypermultiplet in the adjoint representation of $G$, but the four scalars transform as spin 1 plus spin 0 under $\SU(2)_R$ symmetry, according to the decomposition \eqref{tg}. They are neutral under $\cN{=}2$ U$(1)_R$ symmetry.   In general, the single-letter index of a chiral multiplet  is \begin{equation}
\frac{t^{3R_{\cN=1}}v^f -t^{6-3R_{\cN=1}}v^{-f}}{(1-t^3y)(1-t^3y^{-1})}
\end{equation} where $R_{\cN=1}$ is the $\cN{=}1$ R-symmetry used to put the theory on $S^3\times S^1$ as in \cite{Festuccia:2011ws}, and $v$ and $f$ are the fugacity and the charge of a flavor symmetry commuting with the $\cN{=}1$ subalgebra. 
In our case, \begin{equation}
\lambda_{\cN{=}1}=\frac43I_3+\frac13R_{\cN=2},\quad f=-I_3+\frac12 R_{\cN=2}
\end{equation} where $I_3$ is the third component of the generator of $\SU(2)_R$, and $R_{\cN{=}2}$ is the generator of the $\cN{=}2$ U$(1)_R$ symmetry.
Our hypermultiplet behaves as two chiral multiplets with $(I_3,R_{\cN=2})=(1,0)$ and $(0,0)$, resulting in  the single-letter index
\begin{equation}
\frac{1-q}{1+q} \chi_\text{adj}(\bx) = \chi_\text{adj}(\bx) + \frac{2q}{1-q} \chi_\text{adj}(\bx).
\end{equation} The first term in the right hand side is the zero mode, and the second term is the contribution from the non-zero modes. Therefore, the contribution to the index of the non-zero modes is the plethystic exponential of this second term, \begin{equation}
\exp\left[\sum_{n=1}^\infty \frac{2q^{n}}{1-q^{n}}\chi_\text{adj}(\bx^n)\right] = \eta(q,\bx)^{-1}.\label{oneloop}
\end{equation}
Let us denote by $\cH_\text{one-loop}$ the Hilbert space of the non-zero modes whose index is \eqref{oneloop}. 

The total system has the symmetry $G_L\times G_R$, but the non-zero modes $\cH_\text{one-loop}$ only has the symmetry under $G$. This is because we fixed the zero mode at a point on $G$.
The Hilbert space $\cH_\text{total}$ of the total system is then the space of sections of a  vector bundle over $G$ such that the fiber at a point is $\cH_\text{one-loop}$. In mathematical terms, $\cH_\text{total}$ is the representation of $G_L\times G_R$ induced by the representation $\cH_\text{one-loop}$ of $G_\text{diag}\subset G_L\times G_R$: \begin{equation}
\cH_\text{total}=\Ind_{G_\text{diag}}^{G_L\times G_R} \cH_\text{one-loop}.
\end{equation} 
Let us check that the result \eqref{answer} is the same representation of $G_L\times G_R$ as $\cH_\text{total}$. 
This can be done by counting the multiplicity of each irreducible representations. 
The multiplicity of $\lambda_1\times \lambda_2^*$ of $G_L\times G_R$ in $\cH_\text{total}$ is given by the Frobenius reciprocity: \begin{equation}
\int[d\bx] \braket{\lambda_1}{\bx}\eta(q,\bx)^{-1}\braket{\bx}{\lambda_2}.\label{bar}
\end{equation} 
The multiplicity of $\lambda_1\times \lambda_2^*$ of $G_L\times G_R$ in \eqref{answer} is \begin{equation}\begin{aligned}
&\iint [d\bx_L][d\bx_R] \braket{\lambda_1}{\bx_L} Z_{\cA=0}(q,\bx_L,\bx_R)\braket{\bx_R}{\lambda_2} \\
&= 
\iint [d\bx_L][d\bx_R] \braket{\lambda_1}{\bx_L} \eta(q,\bx_L)^{-1/2}\braket{\bx_L}{\bx_R}\eta(q,\bx_R)^{-1/2}\braket{\bx_R}{\lambda_2}\\
&=\int[d\bx] \braket{\lambda_1}{\bx}\eta(q,\bx)^{-1}\braket{\bx}{\lambda_2}
\end{aligned}\end{equation} which is equal to \eqref{bar}.

To find the Hamiltonian of the system, one needs to determine the Berry-phase connection of the bundle $\cH_\text{one-loop}$ over $G$. Then the total Hamiltonian is obtained as the Laplacian acting on the sections of the bundle.  The author has not attempted to perform this analysis. The Berry phase of the supersymmetric quantum system was studied in e.g.~\cite{Pedder:2007ff,Sonner:2008fi}, which might be useful in pursuing this calculation.

\section{Generalization}\label{generalization}
Let $p=t^3y$, $q=t^3y^{-1}$, $\ut=t^4/v$, and let $p\to 0$ keeping $q$ and $\ut$ fixed\footnote{Our $t$ is the $t$ of \cite{Gadde:2009kb}. Our $\ut$ is the $t$ of \cite{Gadde:2011uv}.}.  It was found in \cite{Gadde:2011uv} that the 2d q-deformed Yang-Mills is then further deformed, such that e.g.~the characters $\chi_\lambda(\bx)$ is replaced by Macdonald functions $P_\lambda(q,\ut,\bx)$. The contribution from the gauge modes \eqref{gauge}  is now \begin{equation}
\xi(q,\ut,\bx)=\exp\left[\sum_{n=1}^\infty \frac{-q^{n}-\ut^n}{1-q^{n}}\chi_\text{adj}(\bx^n)\right].
\end{equation} The analysis of the last section can be repeated easily. The non-zero modes of the hypermultiplets \eqref{oneloop} contribute by\begin{equation}
\xi(q,\ut,\bx)^{-1}=\exp\left[\sum_{n=1}^\infty \frac{q^{n}+\ut^n}{1-q^{n}}\chi_\text{adj}(\bx^n)\right],
\end{equation} and then the multiplicity of $\lambda_1\times \lambda_2^*$ in $\cH_\text{total}$  is \begin{equation}
\int[d\bx] \braket{\lambda_1}{\bx}\xi(q,\ut,\bx)^{-1}\braket{\bx}{\lambda_2}. \label{bosh}
\end{equation} The generalization of the formula of $Z_{\cA}$ \eqref{answer} would be \begin{equation}
Z_{\cA}(q,\ut,\bx_L,\bx_R)=\zeta(q,\ut,\bx_L,\bx_R)\sum_\lambda \frac{P_\lambda(q,\ut,\bx_L)  e^{-C_2(\lambda)\cA\beta/\vol S^3} P_{\lambda}(q,\ut,\bx_R^{-1})}{\cN_{\lambda}(q,\ut)}  \label{guess}
\end{equation} where $P_\lambda$ is the Macdonald function, and $\cN_{\lambda}(q,\ut)=\langle P_\lambda,P_\lambda\rangle_{q,\ut}$ is its norm.\footnote{We normalize $P_\lambda$ as in the mathematics literature, not as in \cite{Gadde:2011uv}. So, our $P_\lambda$'s are orthogonal but not orthonormal.}
The  factor $\zeta(q,\ut,\bx_L,\bx_R)$ is determined by the generalization of \eqref{composition} \begin{equation}
\int [d\bx] Z_\cA(q,\ut,\bx_L,\bx)\xi(q,\ut,\bx)Z_{\cA'}(q,\ut,\bx,\bx_R) = Z_{\cA+\cA'}(q,\ut,\bx_L,\bx_R).
\end{equation} A short calculation yields \begin{equation}
\zeta(q,\ut,\bx_L,\bx_R)=\eta(q,\bx_L)^{-1/2}\eta(q,\bx_R)^{-1/2}\prod_{n=0}^\infty\left(\frac{1-\ut q^n}{1-q^{n+1}}\right)^{r}\label{fudge} 
\end{equation} where $r$ is the rank of $G$.

One finds by a straightforward calculation using the orthogonality of Macdonald polynomials  that the multiplicity of $\lambda_1\times \lambda_2^*$ in \eqref{guess}  is equal to the expression \eqref{bosh}.  We can perform a finer decomposition by  counting the multiplicity of $\lambda_1\times \lambda_2^*$ in \eqref{guess} which has the energy $C_2(\lambda)$. 
This is given by \begin{equation}
\iint [d\bx_L] [d\bx_R] \zeta(q,\ut,\bx_L,\bx_R) \chi_{\lambda_1^*}(\bx_L)  
\frac{P_\lambda(q,\ut,\bx_L)P_{\lambda}(q,\ut,\bx_R^{-1})}{\cN_\lambda(q,\ut)}
 \chi_{\lambda_2}(\bx_R).\label{qqq}
\end{equation} This should be a series in $q$ and $\ut$ with \emph{interger} coefficients, and indeed it is, 
because $\cN_\lambda(q,\ut)$ and the expansions of $\eta(q,\ut,\bx)^{-1/2}$, $P_\lambda(q,\ut,\bx)$ in terms of  $\chi_{\lambda'}(\bx)$ are all series with integer coefficients. %\footnote{The last integrality is basically Macdonald's conjecture, which is now proven.}

The 4d theory associated to a once-punctured sphere is also an $\cN=2$ non-linear sigma model \cite{Gaiotto:2011xs}. Its partition function on $S^1\times S^3$ will essentially be the Macdonald function. It seems promising to pursue this line of  ideas. 

%-----------------------------------------------------------------
\section*{Acknowledgements}
%-----------------------------------------------------------------
The author thanks Yasuhiro Yamamoto for discussions.
This work is  supported in part by World Premier International Research Center Initiative
(WPI Initiative),  MEXT, Japan through the Institute for the Physics and Mathematics
of the Universe, the University of Tokyo.

\bibliographystyle{ytphys}
\small\baselineskip=.93\baselineskip
\bibliography{refX}

\providecommand{\href}[2]{#2}\begingroup\raggedright\begin{thebibliography}{10}

\bibitem{Gadde:2009kb}
A.~Gadde, E.~Pomoni, L.~Rastelli, and S.~S. Razamat, ``{S-Duality and 2d
  Topological QFT},'' \href{http://dx.doi.org/10.1007/JHEP03(2010)032}{{\em
  JHEP} {\bfseries 03} (2010) 032},
\href{http://arxiv.org/abs/0910.2225}{{\ttfamily arXiv:0910.2225 [hep-th]}}.
%%CITATION = 0910.2225;%%.

\bibitem{Gadde:2011ik}
A.~Gadde, L.~Rastelli, S.~S. Razamat, and W.~Yan, ``{The 4d Superconformal
  Index from q-Deformed 2d Yang- Mills},''
  \href{http://dx.doi.org/10.1103/PhysRevLett.106.241602}{{\em Phys. Rev.
  Lett.} {\bfseries 106} (2011) 241602},
\href{http://arxiv.org/abs/1104.3850}{{\ttfamily arXiv:1104.3850 [hep-th]}}.
%%CITATION = 1104.3850;%%.

\bibitem{Gadde:2011uv}
A.~Gadde, L.~Rastelli, S.~S. Razamat, and W.~Yan, ``{Gauge Theories and
  Macdonald Polynomials},''
\href{http://arxiv.org/abs/1110.3740}{{\ttfamily arXiv:1110.3740 [hep-th]}}.
%%CITATION = ARXIV:1110.3740;%%.

\bibitem{Gaiotto:2009we}
D.~Gaiotto, ``{${\mathcal{N}}\!=2$ Dualities},''
\href{http://arxiv.org/abs/0904.2715}{{\ttfamily arXiv:0904.2715 [hep-th]}}.
%%CITATION = 0904.2715;%%.

\bibitem{Gaiotto:2011xs}
D.~Gaiotto, G.~W. Moore, and Y.~Tachikawa, ``{On 6d $\cN=$(2,0) Theory
  Compactified on a Riemann Surface with Finite Area},''
\href{http://arxiv.org/abs/1110.2657}{{\ttfamily arXiv:1110.2657 [hep-th]}}.
%%CITATION = 1110.2657;%%.

\bibitem{Benini:2012ui}
F.~Benini and S.~Cremonesi, ``{Partition Functions of $\cN=$(2,2) Gauge
  Theories on S$^2$ and Vortices},''
\href{http://arxiv.org/abs/1206.2356}{{\ttfamily arXiv:1206.2356 [hep-th]}}.
%%CITATION = ARXIV:1206.2356;%%.

\bibitem{Doroud:2012xw}
N.~Doroud, J.~Gomis, B.~Le~Floch, and S.~Lee, ``{Exact Results in $d=2$
  Supersymmetric Gauge Theories},''
\href{http://arxiv.org/abs/1206.2606}{{\ttfamily arXiv:1206.2606 [hep-th]}}.
%%CITATION = ARXIV:1206.2606;%%.

\bibitem{Kapustin:2009kz}
A.~Kapustin, B.~Willett, and I.~Yaakov, ``{Exact Results for Wilson Loops in
  Superconformal Chern-Simons Theories with Matter},'' {\em JHEP} {\bfseries
  1003} (2010) 089,
\href{http://arxiv.org/abs/0909.4559}{{\ttfamily arXiv:0909.4559 [hep-th]}}.
%%CITATION = ARXIV:0909.4559;%%.

\bibitem{Pestun:2007rz}
V.~Pestun, ``{Localization of Gauge Theory on a Four-Sphere and Supersymmetric
  Wilson Loops},''
\href{http://arxiv.org/abs/0712.2824}{{\ttfamily arXiv:0712.2824 [hep-th]}}.
%%CITATION = 0712.2824;%%.

\bibitem{Kallen:2012cs}
J.~K{\"all\'en} and M.~Zabzine, ``{Twisted Supersymmetric 5D Yang-Mills Theory
  and Contact Geometry},''
  \href{http://dx.doi.org/10.1007/JHEP05(2012)125}{{\em JHEP} {\bfseries 1205}
  (2012) 125},
\href{http://arxiv.org/abs/1202.1956}{{\ttfamily arXiv:1202.1956 [hep-th]}}.
%%CITATION = ARXIV:1202.1956;%%.

\bibitem{Hosomichi:2012ek}
K.~Hosomichi, R.-K. Seong, and S.~Terashima, ``{Supersymmetric Gauge Theories
  on the Five-Sphere},''
\href{http://arxiv.org/abs/1203.0371}{{\ttfamily arXiv:1203.0371 [hep-th]}}.
%%CITATION = ARXIV:1203.0371;%%.

\bibitem{Kallen:2012va}
J.~K{\"all\'en}, J.~Qiu, and M.~Zabzine, ``{The Perturbative Partition Function
  of Supersymmetric 5D Yang-Mills Theory with Matter on the Five-Sphere},''
\href{http://arxiv.org/abs/1206.6008}{{\ttfamily arXiv:1206.6008 [hep-th]}}.
%%CITATION = ARXIV:1206.6008;%%.

\bibitem{Kim:2012av}
H.-C. Kim and S.~Kim, ``{M5-Branes from Gauge Theories on the 5-Sphere},''
\href{http://arxiv.org/abs/1206.6339}{{\ttfamily arXiv:1206.6339 [hep-th]}}.
%%CITATION = ARXIV:1206.6339;%%.

\bibitem{Kallen:2012zn}
J.~K{\"all\'en}, J.~Minahan, A.~Nedelin, and M.~Zabzine, ``{N$^3$-Behavior from
  5D Yang-Mills Theory},''
\href{http://arxiv.org/abs/1207.3763}{{\ttfamily arXiv:1207.3763 [hep-th]}}.
%%CITATION = ARXIV:1207.3763;%%.

\bibitem{Alday:2009aq}
L.~F. Alday, D.~Gaiotto, and Y.~Tachikawa, ``{Liouville Correlation Functions
  from Four-Dimensional Gauge Theories},''
  \href{http://dx.doi.org/10.1007/s11005-010-0369-5}{{\em Lett. Math. Phys.}
  {\bfseries 91} (2010) 167--197},
\href{http://arxiv.org/abs/0906.3219}{{\ttfamily arXiv:0906.3219 [hep-th]}}.
%%CITATION = 0906.3219;%%.

\bibitem{Wyllard:2009hg}
N.~Wyllard, ``{$A_{N-1}$ Conformal Toda Field Theory Correlation Functions from
  Conformal ${\mathcal{N}}\!=2$ $SU(N)$ Quiver Gauge Theories},''
  \href{http://dx.doi.org/10.1088/1126-6708/2009/11/002}{{\em JHEP} {\bfseries
  11} (2009) 002},
\href{http://arxiv.org/abs/0907.2189}{{\ttfamily arXiv:0907.2189 [hep-th]}}.
%%CITATION = 0907.2189;%%.

\bibitem{Dimofte:2011ju}
T.~Dimofte, D.~Gaiotto, and S.~Gukov, ``{Gauge Theories Labelled by
  Three-Manifolds},''
\href{http://arxiv.org/abs/1108.4389}{{\ttfamily arXiv:1108.4389 [hep-th]}}.
%%CITATION = ARXIV:1108.4389;%%.

\bibitem{Terashima:2011xe}
Y.~Terashima and M.~Yamazaki, ``{Semiclassical Analysis of the 3d/3d
  Relation},''
\href{http://arxiv.org/abs/1106.3066}{{\ttfamily arXiv:1106.3066 [hep-th]}}.
%%CITATION = ARXIV:1106.3066;%%.

\bibitem{Dimofte:2011py}
T.~Dimofte, D.~Gaiotto, and S.~Gukov, ``{3-Manifolds and 3d Indices},''
\href{http://arxiv.org/abs/1112.5179}{{\ttfamily arXiv:1112.5179 [hep-th]}}.
%%CITATION = ARXIV:1112.5179;%%.

\bibitem{Kawano:2012up}
T.~Kawano and N.~Matsumiya, ``{5d SYM on 3d Sphere and 2d YM},''
\href{http://arxiv.org/abs/1206.5966}{{\ttfamily arXiv:1206.5966 [hep-th]}}.
%%CITATION = ARXIV:1206.5966;%%.

\bibitem{Kim:2012gu}
H.-C. Kim, S.-S. Kim, and K.~Lee, ``{5-Dim Superconformal Index with Enhanced
  $E_n$ Global Symmetry},''
\href{http://arxiv.org/abs/1206.6781}{{\ttfamily arXiv:1206.6781 [hep-th]}}.
%%CITATION = ARXIV:1206.6781;%%.

\bibitem{Terashima:2012ra}
S.~Terashima, ``{On Supersymmetric Gauge Theories on S$^4$ $\times$ S$^1$},''
\href{http://arxiv.org/abs/1207.2163}{{\ttfamily arXiv:1207.2163 [hep-th]}}.
%%CITATION = ARXIV:1207.2163;%%.

\bibitem{Festuccia:2011ws}
G.~Festuccia and N.~Seiberg, ``{Rigid Supersymmetric Theories in Curved
  Superspace},'' \href{http://dx.doi.org/10.1007/JHEP06(2011)114}{{\em JHEP}
  {\bfseries 1106} (2011) 114},
\href{http://arxiv.org/abs/1105.0689}{{\ttfamily arXiv:1105.0689 [hep-th]}}.
%%CITATION = ARXIV:1105.0689;%%.

\bibitem{Hama:2010av}
N.~Hama, K.~Hosomichi, and S.~Lee, ``{Notes on SUSY Gauge Theories on
  Three-Sphere},'' \href{http://dx.doi.org/10.1007/JHEP03(2011)127}{{\em JHEP}
  {\bfseries 1103} (2011) 127},
\href{http://arxiv.org/abs/1012.3512}{{\ttfamily arXiv:1012.3512 [hep-th]}}.
%%CITATION = ARXIV:1012.3512;%%.

\bibitem{Hama:2011ea}
N.~Hama, K.~Hosomichi, and S.~Lee, ``{SUSY Gauge Theories on Squashed
  Three-Spheres},'' {\em JHEP} {\bfseries 1105} (2011) 014,
\href{http://arxiv.org/abs/1102.4716}{{\ttfamily arXiv:1102.4716 [hep-th]}}.
%%CITATION = ARXIV:1102.4716;%%.

\bibitem{Imamura:2011wg}
Y.~Imamura and D.~Yokoyama, ``{${\mathcal{N}}\!=2$ Supersymmetric Theories on
  Squashed Three-Sphere},'' {\em Phys.Rev.} {\bfseries D85} (2012) 025015,
\href{http://arxiv.org/abs/1109.4734}{{\ttfamily arXiv:1109.4734 [hep-th]}}.
%%CITATION = ARXIV:1109.4734;%%.

\bibitem{Hama:2012bg}
N.~Hama and K.~Hosomichi, ``{Seiberg-Witten Theories on Ellipsoids},''
\href{http://arxiv.org/abs/1206.6359}{{\ttfamily arXiv:1206.6359 [hep-th]}}.
%%CITATION = ARXIV:1206.6359;%%.

\bibitem{Benini:2011nc}
F.~Benini, T.~Nishioka, and M.~Yamazaki, ``{4d Index to 3d Index and 2d
  TQFT},''
\href{http://arxiv.org/abs/1109.0283}{{\ttfamily arXiv:1109.0283 [hep-th]}}.
%%CITATION = ARXIV:1109.0283;%%.

\bibitem{Gaiotto:2008sa}
D.~Gaiotto and E.~Witten, ``{Supersymmetric Boundary Conditions in
  ${\mathcal{N}}\!=4$ Super Yang-Mills Theory},''
\href{http://arxiv.org/abs/0804.2902}{{\ttfamily arXiv:0804.2902 [hep-th]}}.
%%CITATION = 0804.2902;%%.

\bibitem{Cordes:1994fc}
S.~Cordes, G.~W. Moore, and S.~Ramgoolam, ``{Lectures on 2-D Yang-Mills Theory,
  Equivariant Cohomology and Topological Field Theories},''
  \href{http://dx.doi.org/10.1016/0920-5632(95)00434-B}{{\em
  Nucl.Phys.Proc.Suppl.} {\bfseries 41} (1995) 184--244},
\href{http://arxiv.org/abs/hep-th/9411210}{{\ttfamily arXiv:hep-th/9411210
  [hep-th]}}.
%%CITATION = HEP-TH/9411210;%%.

\bibitem{Pedder:2007ff}
C.~Pedder, J.~Sonner, and D.~Tong, ``{The Geometric Phase in Supersymmetric
  Quantum Mechanics},''
  \href{http://dx.doi.org/10.1103/PhysRevD.77.025009}{{\em Phys.Rev.}
  {\bfseries D77} (2008) 025009},
\href{http://arxiv.org/abs/0709.0731}{{\ttfamily arXiv:0709.0731 [hep-th]}}.
%%CITATION = ARXIV:0709.0731;%%.

\bibitem{Sonner:2008fi}
J.~Sonner and D.~Tong, ``{Berry Phase and Supersymmetry},''
  \href{http://dx.doi.org/10.1088/1126-6708/2009/01/063}{{\em JHEP} {\bfseries
  0901} (2009) 063},
\href{http://arxiv.org/abs/0810.1280}{{\ttfamily arXiv:0810.1280 [hep-th]}}.
%%CITATION = ARXIV:0810.1280;%%.

\end{thebibliography}\endgroup

\end{document}